\documentclass[11pt]{article}

\usepackage{amsmath,amssymb,amsthm,epsf,epsfig}
\usepackage{amsmath,amssymb}

\begin{document}
\title{Exceptional orthogonal polynomials and exactly solvable potentials in position-dependent-mass Schroedinger
 Hamiltonians}
\author{B.Midya\thanks{E-mail: bikash.midya@gmail.com} ~and ~B. Roy
\thanks{E-mail : barnana@isical.ac.in~~~~~Fax: +9133 2577-3026}
 ~ \\ Physics \& Applied
Mathematics Unit\\ Indian Statistical Institute \\ Calcutta
700108\\ India}
 \maketitle

\vspace*{.75 cm}

\centerline{\bf Abstract} Some exactly solvable potentials in the position dependent mass background are generated whose bound states are given in terms of Laguerre- or
Jacobi-type $X_1$ exceptional orthogonal polynomials. These potentials are shown to be shape invariant and isospectral to the potentials whose bound state solutions
involve classical Laguerre or Jacobi polynomials.
 \vspace{1 cm}


 \vspace{.2 cm}


\section{Introduction}In recent years, quantum mechanical systems with a position dependent
 mass have attracted a lot of interest due to their relevance in describing the physics
 of many microstructures of current interest, such as compositionally graded crystals {\cite{r1}},
 quantum dots {\cite{r2}}, $^3$He clusters {\cite{r3}}, metal clusters {\cite{r4}} etc. The concept of position
  dependent mass comes from the effective mass approximation {\cite{r41}}-{\cite{r43}} which is an useful tool for
    studying the motion of carrier electrons in pure crystals and also for the virtual-crystal approximation in the
  treatment of homogeneous alloys (where the actual potential is approximated by a periodic potential) as well as in
  graded mixed semiconductors (where the potential is not periodic). The attention to the effective mass approach stems
  from the extraordinary development in crystallographic growth techniques which allow the production of non uniform semiconductor
  specimen with abrupt heterojunctions. In these mesoscopic materials, the effective mass of the charge carriers are
  position dependent.  Consequently the study of the effective mass Schr\"{o}dinger equation becomes relevant for deeper
   understanding of the non-trivial quantum effects observed on these nanostructures. The position dependent (effective)
   mass is also used in the construction of pseudo-potentials which have a significant computational advantage
   in quantum Monte Carlo method{\cite{r5,r40}}. This large variety of applications explains the
growing interest in constructing solvable cases of effective mass
Schr\"{o}dinger equation. Recently, several of such cases were
obtained by means of point canonical transformation (PCT), Lie
algebraic techniques and supersymmetric quantum mechanical methods
{\cite{r7}-{\cite{r25}}.

In this letter, our objective is to construct new exactly solvable
potentials in a position dependent effective mass background whose
bound state wavefunctions can be written in terms of recently
 introduced {\cite{r28,r29} Laguerre or Jacobi type $X_1$ exceptional orthogonal polynomials.
 (It must be mentioned here that similar study has been done by Quesne \cite{r27} in the constant mass scenario)
 Subsequently, we shall show that these exactly solvable potentials are shape invariant by using supersymmetric quantum
 mechanics (SUSYQM) technique in position dependent mass case. SUSYQM allows us to obtain isospectral partner potentials of these exactly solvable potentials.
  The motivation for doing this,
 apart from enlarging the class of exactly solvable potentials for position dependent mass Schr\"{o}dinger equation,
  comes from the fact that in different areas of possible applications of low dimensional structures e.g. quantum well,
   quantum dot, there is need to have energy spectrum which is predetermined. Specifically, in the quantum well profile
   optimization isospectral potentials are generated through supersymmetric quantum mechanics. These are necessary because
   a particular effect such as intersubband optical transitions in a quantum well, may be grossly enhanced by achieving the resonance conditions i.e.
   appropriate spacings between the most relevant states and also by tailoring the wave functions so that the (combinations of) matrix elements relevant
   for this particular effect are maximized \cite{r56}. This is particularly important for higher order nonlinear processes.
    Optimization of simple stepwise constant profiled quantum wells has been considered quite some time ago \cite{r57}, while
    the optimization of continuously graded structure requires more sophisticated techniques like supersymmetric quantum mechanics and inverse
    spectral theory \cite{r58}.

The organization of this article
 is as follows. In section 2. we have  generated two new exactly solvable
potentials for one dimensional position dependent mass Schrodinger
equation whose wavefunctions involve Laguerre or Jacobi-type $X_1$
extended polynomials via point canonical transformation
approach(PCT) \cite{r33,r34}. In section 3. we have shown that the
new potentials are shape invariant in the context of
supersymmetric quantum mechanics in the position dependent mass
background. Section 4. is kept for summary and discussions.

\section{\bf Generation of new potentials via PCT}
The general position dependent mass (time independent)
Schr\"{o}dinger equation, initially proposed by von Roos
{\cite{r6}} in terms of three ambiguity parameters $r$, $s$, $t$
such that $r+s+t = -1$ is given by
\begin{equation}
H_{eff}\psi(x)=E \psi(x)~,~~~~~H_{eff}=-\frac{d}{dx}
\frac{1}{M(x)} \frac{d}{dx} +V_{eff}\label{e1}
\end{equation}
where the effective potential
\begin{equation}
V_{eff}=V(x)+\frac{s+1}{2}\frac{M''}{M^2}-[r(r+s+1)+s+1]\frac{M'^2}{M^3}\label{e2}
\end{equation}
depends on some mass term $M(x)$. Here prime denotes
differentiation with respect to $x$. $M(x)$ is the dimensionless
form of the mass function $m(x) = m_0 M(x)$ and we have taken
$\hbar = 2m_0 = 1$.

 Let us find the solution of (\ref{e1}) of the form
 \begin{equation}
 \psi(x)=f(x)F\left(g(x)\right)\label{e3}
 \end{equation}
 where $f(x), g(x)$ are two function of $x$ to be determined and
 $F(g)$ satisfies second order differential equation of
 the type
 \begin{equation}
 \frac{d^2F}{dg^2}+Q(g) \frac{dF}{dg}+R(g) F=0 \label{e4}
 \end{equation}\\

 Now in order to be physically acceptable, $\psi(x)$ has to satisfy the following two conditions : \\
 (i) Like quantum systems with constant mass it will be square integrable over domain of definition {\bf{D}} of $M(x)$
 and
 $\psi(x)$ i.e.,$$\int_D |\psi_n(x)|^2 dx<\infty$$\\
 (ii) The Hermiticity of the Hamiltonian in the Hilbert
  space spanned by the eigenfunctions of the potential $V(x)$ is ensured by the following extra
 condition {\cite{r19}}
 $$\frac{|\psi_n(x)|^2}{\sqrt{M(x)}}\rightarrow 0$$ at the end points of the interval where $V(x)$ and $\psi_n(x)$ are
  defined. This condition imposes an additional restriction whenever the mass function $M(x)$ vanishes at any one or both
  the end points of $\mathcal {D}$.

 Substituting $\psi(x)$ from equation (\ref{e3}) in  equation (\ref{e1}) leads to
 \begin{equation}
 \frac{d^2F}{dg^2}+\left(\frac{g''}{g'^2}+\frac{2f'}{fg'}-\frac{M'}{Mg'}\right)
 \frac{dF}{dg}+\left(\frac{f''}{fg'^2}+(E-V_{eff})\frac{M}{g'^2}-\frac{M'f'}{Mfg'^2}\right)F=0\label{e5}
 \end{equation}
Comparing equations (\ref{e4}) and (\ref{e5}) we get
\begin{equation}
\begin{array}{lcl}
Q(g(x)) &=& \frac{g''}{g'^2}+\frac{2f'}{fg'}-\frac{M'}{Mg'}\\
R(g(x)) &=& \frac{f''}{fg'^2}+(E-V_{eff})\frac{M}{g'^2}-\frac{M'f'}{Mfg'^2}\label{e6}
\end{array}
\end{equation}
The first of Eqns. (\ref{e6}) gives
\begin{equation}
f(x)\propto \sqrt{\frac{M}{g'}}
~~~~exp\left(\frac{1}{2}\int^{g(x)}Q(t)dt\right)\label{e7}
\end{equation}
while the latter leads to the equation
\begin{equation}
E-V_{eff}=\frac{g'''}{2Mg'}-\frac{3}{4M}\left(\frac{g''}{g'}\right)^2+\frac{g'^2}{M}
\left(R-\frac{1}{2}\frac{dQ}{dg}-\frac{Q^2}{4}\right)-\frac{M''}{2M^2}+\frac{3M'^2}{4M^3}
\label{e8}
\end{equation}
For equation (\ref{e8}) to be satisfied, we need to find some functions $M(x)$, $g(x)$
 ensuring the presence of a constant term on the right hand side of equation (\ref{e8}) to
  compensate $E$ on its left hand side and giving rise to an effective potential $V_{eff}(x)$ with well
   behaved wavefunctions.

Now in PCT approach there are many options for choosing $M(x)$,
for example $M(x) = \lambda g'^2(x)$ {\cite{r10,r22}}, $M =
\lambda g'(x)$ , $M = \displaystyle \frac{\lambda}{g'(x)}$
{\cite{r26}} , $\lambda$
 being a constant. Here we choose $M(x)=\lambda g'(x)$ so that equation (\ref{e8}) reduces to
\begin{equation}
E-V_{eff}=\frac{g'}{\lambda}\left(R-\frac{1}{2}\frac{dQ}{dg}-\frac{Q^2}{4}\right)\label{e9}
\end{equation}\\
{\bf{Example 1 :}}\\
First we consider $F_n
(g)\propto\hat{L}_n^{(\alpha)}$, where
$\hat{L}_n^{(\alpha)},~n=1,2,3,...,~ \alpha>0$ is the Laguerre
type $X_1$ polynomials {\cite{r29}}. For this polynomial
$$Q(g)=-\frac{(g-\alpha)(g+\alpha+1)}{g(g+\alpha)}~~~~~~~~ R(g)=\frac{n-2}{g}+\frac{2}{g+\alpha}.$$
For these values of $Q(g)$ and $R(g)$ equation (9) becomes
\begin{equation}
E-V_{eff}=\frac{g'}{\lambda g}\left(\frac{2\alpha
n+\alpha^2-\alpha+2}{2\alpha}\right)-\frac{g'}{g^2}
\left(\frac{\alpha^2-1}{4\lambda}\right)-\frac{2g'}{\lambda(g+\alpha)^2}-\frac{g'}{\lambda
\alpha(g+\alpha)}-\frac{g'}{4\lambda} \label{e10}
\end{equation}
Taking $\frac{g'}{\lambda g}=C$,where $C$ is a constant, a
constant term can be created on the right-hand side of the above
equation which will correspond to $E$ on the left-hand side. But
$C$ must be restricted to positive
 values in order to get increasing energy eigenvalues for successive $n$ values.
  The solution of the above-mentioned first order differential equation for $g(x)$ leading to
  a positive mass function reads
\begin{equation}
g(x)=e^{-bx} ~~~~,~~~~~~~~~~ M(x)=e^{-bx}~,~~~~~~~
-\infty<x<\infty\label{e11}
\end{equation}
where $C\lambda=-b,b>0.$ This exponentially behaved mass function
is often used in the study of confined energy states for carriers
in semiconductor quantum well \cite{r21,r51}. This mass function
was also found to be useful for understanding transport properties
through semiconductor heterostructures (produced by the recent
crystallographic growth techniques) which is indispensable for the
prediction of the performances of these samples . In this context,
this mass function has been used to compute transmission
probabilities for scattering in abrupt heterostructures \cite{r52}
which may be useful in the
design of semiconductor devices \cite{r55}.\\
Now equations (\ref{e7}),(\ref{e10}), and (\ref{e11}) give
$$
E=\frac{b^2}{4}
\left(4n+\frac{4}{\alpha}+2\alpha-2\right)+\bar{V}_0$$
\begin{equation}
V_{eff}=\left[\frac{b^2}{4}\left((\alpha^2-1)e^{bx}+e^{-bx}\right)\right]+\frac{b^2}{4}\left(\frac{4}{\alpha(1+\alpha
e^{bx})}+\frac{8e^{bx}}{(1+\alpha
e^{bx})^2}\right)+\bar{V}_0\label{e12}
\end{equation}
where $\bar {V_0}$ is a constant. Thus we obtain
energy eigenvalues and eigenfunctions as
\begin{equation}
E_m=b^2
\left(m+\frac{\alpha+1}{2}\right)+\frac{b^2}{\alpha}+\bar{V}_0\label{e43}
\end{equation}
\begin{equation}
\psi_m(x)={\mathcal{N}_m}
~~\frac{exp\left[-\frac{1}{2}\{(\alpha+1)bx+e^{-bx}\}\right]}{\alpha+e^{-bx}}~~\hat{L}_{m+1}^{(\alpha)}
(e^{-bx}),~~~~~m=0,1,2...\label{e14}
\end{equation}
where we have reset quantum number $n = m+1$ and $\mathcal{N}_m$
is the normalization constant given by {\cite{r28}}
$$\mathcal{N}_m=\left(\frac{b~
m!}{(m+\alpha+1) \Gamma(m+\alpha)}\right)^{1/2}$$\\
It is to be noted that the square integrability condition(i)
stated earlier does not impose any additional restriction on
$\alpha$ but for condition(ii) to be satisfied, $\alpha$ should be
greater than $-\frac{1}{2}$. In figure 1 we have plotted the
 mass function $M(x)$ given in (\ref{e11}), potential $V_{eff}$ given in (\ref{e12}) and square of first two bound
 state wavefunctions.

\begin{figure}[h]
\epsfxsize=4.5 in \epsfysize=3 in \centerline{\epsfbox{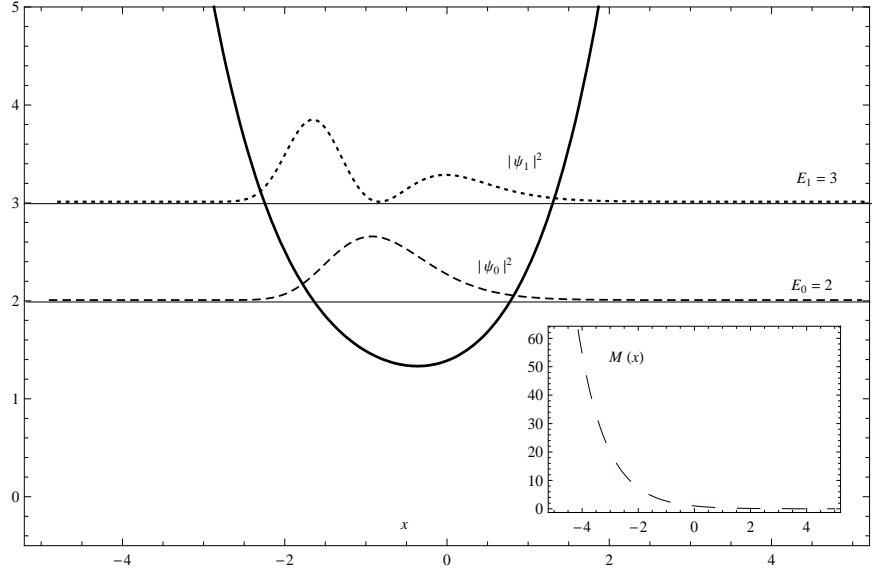}}
\label*{}\caption{ Plot of the potential $V_{eff}$ (solid line)
given in equation (\ref{e12}), square of its first two bound state
wave functions $|\psi_o(x)|^2$ (dashed line) and $|\psi_1(x)|^2$
(dotted line), for the mass function $M(x)$ (long dashed line)
given in equation (\ref{e11}). We have considered here $b=1,
\alpha=2.$}
\end{figure}

{\bf{ Example 2 :}}\\ Next we choose $F_n(g)\propto
\hat{P}_n^{(\alpha,\beta)}(x)$, where
$\hat{P}_n^{(\alpha,\beta)}$, $n=1,2,3,...\cdots$,
$\alpha,\beta>-1,~\alpha\neq\beta$ is the Jacobi type $X_1$
polynomial. For this polynomial, we have
\begin{equation}
\begin{array}{lcl}
Q(g)&=&-\frac{(\alpha+\beta+2)g-(\beta-\alpha)}{1-g^2}-\frac{2(\beta-\alpha)}{(\beta-\alpha)g-(\beta+\alpha)}\\
R(g)&=&-\frac{(\beta-\alpha)g-(n-1)(n+\alpha+\beta)}{1-g^2}-\frac{(\beta-\alpha)^2}{(\beta-\alpha)g-(\beta+\alpha)}
\end{array}
\end{equation}
For these $Q(g)$ and $R(g)$ equation (\ref{e8}) becomes
\begin{equation}
E-V_{eff}=\frac{g'}{\lambda}\left[\frac{A_1g+A_2}{1-g^2}
+\frac{A_3g+A_4}{(1-g^2)^2}+\frac{A_5}{(\beta-\alpha)g-(\beta+\alpha)
}+\frac{A_6}{[(\beta-\alpha)g-(\beta+\alpha)]^2}\right]\label{e42}
\end{equation}
where
$$A_1=\frac{\beta^2-\alpha^2}{2\alpha \beta}~,~~~~A_2=n^2+(\beta+\alpha-1)n+
\frac{1}{4}\{(\beta+\alpha)^2-2(\beta+\alpha)-4\}+\frac{\beta^2+\alpha^2}{2\alpha\beta}$$
$$A_3=\frac{\beta^2-\alpha^2}{2}~,~~~~A_4=-\frac{\beta^2+\alpha^2-2}{2}~,
~~~~~A_5=\frac{(\beta+\alpha)(\beta-\alpha)^2}{2\alpha \beta}
~,~~~A_6=-2(\beta-\alpha)^2$$ To generate a constant term on the
right-hand side, we suppose  $\frac{g'}{\lambda(1-g^2)}=C_1>0$,
where $C_1$ is a constant. Consequently we obtain
\begin{equation}
g(x)=tanh(ax)~~,~~~~~~~~~~M(x)=sech^2(ax)~,~~~~~~~-\infty<x<\infty.\label{e13}
\end{equation}
where $C_1\lambda=a,~ a>0.$ This asymptotically vanishing mass
function depicts a solitonic profile \cite{r53}. Recently, this
mass profile has been used in position dependent mass Hamiltonians
of Zhu-Kroener \cite{r54} and BenDaniel-Duke \cite{r45} type and
interesting connection was shown \cite{r53} between the discrete
eigenvalues of such Hamiltonians and the stationary 1-soliton and
2-soliton solutions of the Korteweg-de-Vries (kdV) equation that
match with the mass function up to a constant of proportionality.
Also, in dealing with position dependent mass models controlled by
a $sech^2$-mass profile, it was demonstrated \cite{r60} that in
the framework of a first order intertwining relationship, such a
mass environment generates an infinite sequence of bound states
for the conventional free-particle
problem.\\
Now using (\ref{e7}),(\ref{e42}), (\ref{e13}) and setting quantum
number $n=m+1$, we obtain the new potential , energy eigenvalues
and corresponding bound state wavefunctions as
\begin{equation}
V_{eff}=\left[\frac{a^2}{4}\left((\alpha^2-1)e^{2ax}+
(\beta^2-1)e^{-2ax}\right)\right]+\frac{a^2}{4}\left(\frac{4(\alpha-\beta)(\alpha-3\beta)}{\alpha(\beta+\alpha
e^{2ax})}-\frac{8\beta(\alpha-\beta)^2}{\alpha(\beta+\alpha
e^{2ax})^2}\right)+\hat{V_0}\label{e15}
\end{equation}
\begin{equation}
E_m=a^2\left(m+\frac{\alpha+\beta}{2}\right)\left(m+\frac{\alpha+\beta+2}{2}\right)
+a^2\left(\frac{\beta}{\alpha}-\frac{\alpha^2+\beta^2-2}{4}\right)+\hat{V}_0
\label{e16}
\end{equation}
\begin{equation}
\psi_m(x)=\mathcal{N}_m~~\frac{(1-tanh(ax))^{\frac{\alpha+1}{2}}
(1+tanh(ax))^{\frac{\beta+1}{2}}}{\alpha+\beta+(\alpha-\beta)tanh(ax)}~~
\hat{P}_{m+1}^{(\alpha,\beta)}(tanh(ax))~~~~~~~~~~m=0,1,2,...\label{e17}
\end{equation}\\
where $\hat{V}_0$ is a constant and the normalization constant $\mathcal{N}_m$ is given by
{\cite{r28}} ,
$$\mathcal{N}_m=\left(\frac{a(\alpha-\beta)^2~ m!
(2m+\alpha+\beta+1) \Gamma(m+\alpha+\beta+1)}{2^{\alpha+\beta-1}
(m+\alpha+1)(m+\beta+1)\Gamma(m+\alpha)\Gamma(m+\beta)}\right)^{1/2}$$
An additional restriction $\alpha,\beta>-1/2$ is to be imposed to
satisfy condition(ii) stated before whereas the square
integrability condition does
  not require any extra restriction on the parameters. In figure 2 we have plotted the mass function given in (\ref{e13}) and
  potential $V_{eff}$ given in (\ref{e15}) and square of its first two bound state wavefunctions.
\begin{figure}[h]
\epsfxsize=4.5 in \epsfysize=2.75 in
\centerline{\epsfbox{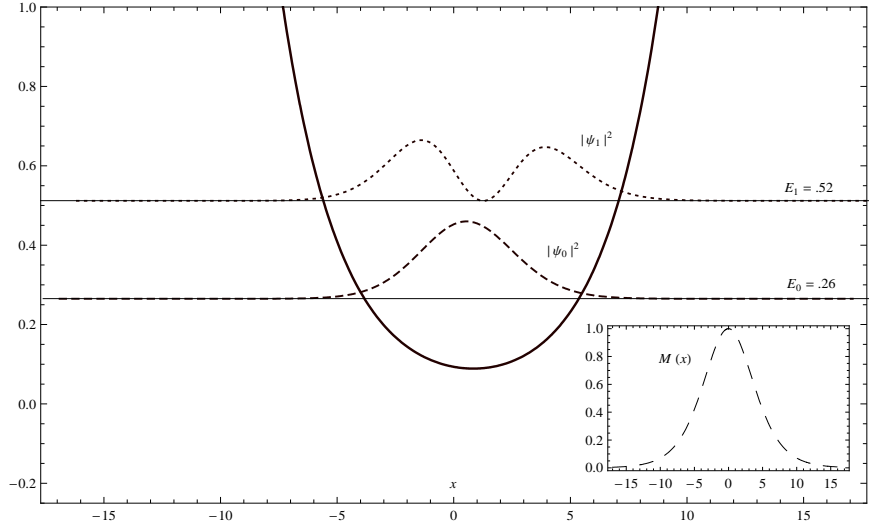}} \label*{}\caption{ Plot of the
potential $V_{eff}$ (solid line) given in equation (\ref{e15}),
square of its first two bound state wave functions $|\psi_o(x)|^2$
(dashed line) and $|\psi_1(x)|^2$ (dotted line), for the mass
function $M(x)$ (long dashed line) given in equation (\ref{e13}).
We have considered here $a=.2, \alpha=2, \beta=2.5.$}
\end{figure}

It should be mentioned here that for the suitable choices of the
constants $\bar{V}_0$ and $\hat{V}_0$, the expressions within the
square bracket of the obtained potentials given in (\ref{e12}) and
(\ref{e15}), coincides with previously obtained potentials in
{\cite{r26} using classical Laguerre or Jacobi polynomial. For the
same choices of $\bar{V}_0$ and $\hat{V}_0$, the potentials
obtained here also become isospectral with the previously obtained
potentials in {\cite{r26}}.

\section{Supersymmetric quantum mechanics approach}
Defining
$$A\psi=\frac{1}{\sqrt{M}}\frac{d\psi}{dx}+B\psi
 ~~~\mbox{and}~~~A^\dag
 \psi=-\frac{d}{dx}\left(\frac{\psi}{\sqrt{M}}\right)+B\psi$$
where $B(x) =-\frac{1}{\sqrt{M}}\frac{\psi_0^{'}}{\psi_0}$ is the
superpotential, the Hamiltonian of equation (\ref{e1}), can be
factorized in the following way
\begin{equation}
H_{eff} = A^{\dagger}A =
-\frac{d}{dx}\left(\frac{1}{M(x)}\right)\frac{d}{dx}+V_{eff}
\end{equation}
It's supersymmetric partner Hamiltonian is given by
\begin{equation}
H_{1,eff} = AA^{\dagger}
=-\frac{d}{dx}\left(\frac{1}{M(x)}\right)\frac{d}{dx}+V_{1,eff}
\end{equation}
where $V_{eff} $ and $V_{1,eff}$  are the supersymmetric partner
potentials  given by
\begin{equation}\begin{array}{ll}
\displaystyle
 V_{eff}=-\left(\frac{B}{\sqrt{M}}\right)'+B^2\\
 \displaystyle
  V_{1,eff}=V_{eff}+\frac{2B'}{\sqrt{M}}-\left(\frac{1}{\sqrt{M}}\right)\left(\frac{1}{\sqrt{M}}\right)''
\end{array}\label{e44}
 \end{equation}
 These two potentials will be called shape invariant if they
 satisfy the condition {\cite{r13,r31}},
\begin{equation}
 V_{1,eff}(x,a_1)=V_{eff}(x,a_2)+R(a_1)\label{e50}
 \end{equation}
 where $a_1$ is a set of parameters
 , $a_2$ is some function of $a_1$ and  $R(a_1)$ is independent of $x$.
In case of unbroken supersymmetry , the energy spectrum and wavefunctions
of two such shape invariant effective mass potentials are related
by {\cite{r13},
\begin{equation}
E^{(eff)}_0=0~,~~~~~E_n^{(1,eff)}=E_{n+1}^{(eff)}~,n=0,1,2,...\label{e18}
\end{equation}
 and
\begin{equation}
 \psi_n^{(1,eff)}=\frac{A\psi_{n+1}^{(eff)}}{\sqrt{E_{n+1}^{(eff)}}}~,~~~~
 \psi_{n+1}^{(eff)}=\frac{A^\dag
 \psi_{n}^{(1,eff)}}{\sqrt{E_n^{(1,eff)}}}~~~~,n=0,1,2...\label{e19}
 \end{equation}

 Now the superpotential $B(x)$ for the potentials obtained in the Example 1 and Example 2, can be written as
 \begin{equation}B(x)=\frac{b}{2}\left[(\alpha+1)e^{\frac{bx}{2}}-e^{-\frac{bx}{2}}\right]
 -\frac{b~e^{\frac{3}{2}bx}}{\alpha(\alpha+1)e^{2bx}+(2\alpha+1)e^{bx}+1}\label{e46}
 \end{equation}
 and
 \begin{equation}
\begin{array}{ll}
\displaystyle
 B(x)=\frac{a}{2}\left[(\alpha-\beta)cosh(ax)+(\alpha+\beta+2)sinh(ax)\right]\\
 \displaystyle ~~~~~~~~~~~~+\frac{2a(\alpha-\beta)}{\left[(\alpha+\beta)cosh(ax)
 +(\alpha-\beta)sinh(ax)\right]\left[(\alpha+\beta+2)cosh(ax)+(\alpha-\beta)sinh(ax)\right]}
 \end{array}\label{e47}
 \end{equation}
 respectively.

  Now using (\ref{e44}) and (\ref{e46}) we obtain
  \begin{equation}
 V_{eff}=\frac{b^2}{4}\left[(\alpha^2-1)e^{bx}+e^{-bx}+\frac{4}{\alpha(1+\alpha
e^{bx})}+\frac{8e^{bx}}{(1+\alpha
e^{bx})^2}\right]-b^2\left(\frac{\alpha+1}{2}+\frac{1}{\alpha}\right)\label{e52}
\end{equation}
\begin{equation}
\displaystyle V_{1,eff}=
\frac{b^2}{4}\left[\alpha(\alpha+2)e^{bx}+e^{-bx}+\frac{4}{(\alpha+1)(1+(\alpha+1)
e^{bx})}+\frac{8e^{bx}}{(1+(\alpha+1)
e^{bx})^2}\right]-\frac{b^2}{2}\left(\frac{\alpha^2+\alpha+2}{\alpha+1}\right)
  \label{e48}
  \end{equation}
Also using (\ref{e44}) and (\ref{e47}) we obtain
\begin{equation}\begin{array}{ll}
\displaystyle
 V_{eff}=\frac{a^2}{4}\left[(\alpha^2-1)e^{2ax}+
(\beta^2-1)e^{-2ax}+\frac{4(\alpha-\beta)(\alpha-3\beta)}{\alpha(\beta+\alpha
e^{2ax})}-\frac{8\beta(\alpha-\beta)^2}{\alpha(\beta+\alpha
e^{2ax})^2}\right]\\
\displaystyle ~~~~~~~~~~~~~~~~ -\frac{a^2}{4}\left(2\alpha
\beta+2\alpha+2\beta+2+\frac{4\beta}{\alpha}\right)\label{e53}
\end{array}
\end{equation}
\begin{equation}\begin{array}{ll}
 \displaystyle
V_{1,eff}=\frac{a^2}{4}\left[(\alpha(\alpha+2))e^{2ax}+
(\beta(\beta+2))e^{-2ax}+\frac{4(\alpha-\beta)(\alpha-3\beta-2)}{(\alpha+1)\left((\alpha+1)
e^{2ax}+\beta+1\right)}\right]\\
\displaystyle
~~~~~~~~~~~~~~~-\frac{a^2}{4}\left(\frac{8(\beta+1)(\alpha-\beta)^2}{(\alpha+1)\left((\alpha+1)
e^{2ax}+\beta+1\right)^2}+2\alpha
\beta+\frac{4(\beta+1)}{\alpha+1}\right)
\end{array}\label{e49}
\end{equation}

The potentials $(V_{eff})$ obtained in (\ref{e52}) and (\ref{e53})
are same with the potentials (\ref{e12}) and (\ref{e15}) for
$\bar{V_0}=-b^2(\frac{\alpha+1}{2}+\frac{1}{\alpha})$ and
$\hat{V}_0=-\frac{a^2}{4}\left(2\alpha
\beta+2\alpha+2\beta+2+\frac{4\beta}{\alpha}\right)$ respectively.

From the equation (\ref{e52}) and (\ref{e48}) we observe that the
the potential $V_{eff}$ and it's supersymmetric partner potentials
$V_{1,eff}$ satisfy the following relation
\begin{equation}
V_{1,eff}(x,\alpha)=V_{eff} (x,\alpha+1)+b^2\label{e54}
\end{equation}
Also from equation (\ref{e53}) and (\ref{e49}) it is clear that
the potentials $V_{eff}$ and $V_{1,eff}$ are related by
\begin{equation}V_{1,eff}(x,\alpha,\beta)=V_{eff}
(x,\alpha+1,\beta+1)+a^2(\alpha+\beta+2)\label{e55}
\end{equation}
From the above two relations we observe that  the potentials
obtained in Example 1 and Example 2 satisfy the condition
 (\ref{e50}). So we conclude that the new potentials are shape
 invariant.\\
Now using (68), (73) of {\cite{r29}} and  (\ref{e19}) we have
derived the
 eigenstates of the partner potential of the potential (\ref{e12}) , as
\begin{equation}
 \psi_m^{(1,eff)}(x)\propto \frac{Exp[-\frac{1}{2}\{(\alpha+1)bx+e^{-bx}\}]}{(\alpha+1)e^{\frac{bx}{2}}
 +e^{-\frac{bx}{2}}}~\hat{L}_{m+1}^{(\alpha+1)}(e^{-bx})~~,m=0,1,2...
 \end{equation}
 and using (42),(47) of {\cite{r29}} and (\ref{e19}) we obtain the eigenstates of
 the partner potential of the potential (\ref{e15}) as
 \begin{equation}\psi_{m}^{(1,eff)}(x)\propto \frac{\left(1+tanh(ax)\right)^{\frac{1+\beta}{2}}
 \left(1-tanh(ax)\right)^{\frac{1+\alpha}{2}}}{(\alpha-\beta)sinh(ax)-(\alpha+\beta+2)cosh(ax)}~~
 \hat{P}_{m+1}^{(\alpha+1,\beta+1)}(tanh(ax))~~,~~~~~~~m=0,1,2...\end{equation}

\section{Summary and Outlook}
We have obtained exactly solvable potentials for position
dependent (effective) mass Schr\"{o}dinger equation whose bound
state solutions are given in terms of Laguerre or Jacobi type
$X_1$ exceptional orthogonal polynomials. As mentioned earlier,
the obtained potentials are the generalizations (by some rational
functions) of the previously obtained potentials {\cite {r26}}
whose bound state solutions involve classical Laguerre or Jacobi
orthogonal polynomials. The method discussed here can be used for
other choices of the function $g(x)$ in order to generate other
type of exactly and quasi exactly solvable  potentials for the one
dimensional Schr\"{o}dinger equation with position dependent mass.
We have shown that these potentials are shape invariant and are
isospectral to the previously obtained potentials in position
dependent mass background whose solutions are given in terms of
classical Laguerre  or Jacobi type orthogional polynomials. Though
the origin of such isospectrality in the constant mass scenario
 has recently been shown in \cite{r35} , it will be worthwhile to study the  origin of such isospectrality in
 the position dependent mass background. As to the possible physical applications of our obtained potentials in the position
 dependent mass background, let us mention that it will be interesting to use supersymmetric quantum mechanics  to generate
 isospectral potentials that depend on a specified number of scalar parameters by multiple deletion and restoring of some levels of the
 original potential in the position dependent mass scenario so as to make them suitable for multiparameter
 optimization of optical nonlinearities in semiconductor quantum
 wells. This particular problem was dealt in  ref \cite{r59},
 taking the constant effective mass $m^*$.

 \vspace{.5cm}

{\bf Acknowledgement}: It is a pleasure to thank Rajkumar
Roychoudhury for many valuable comments and suggestions. We also
thank the referees for their valuable suggestions towards improving
the manuscript.

\end{document}